\documentclass[twocolumn,notitlepage,pra,superscriptaddress]{revtex4-2}
\usepackage{graphicx}
\usepackage{braket}

\usepackage{amsmath,amsfonts,amssymb,empheq}
\usepackage{comment,float}
\usepackage{tabularx}

\usepackage[colorlinks=true,linkcolor=blue,urlcolor=magenta,citecolor=magenta]{hyperref}
\usepackage[normalem]{ulem}
\usepackage{dsfont}
\usepackage{braket}
\usepackage{rotating}
\usepackage{subfig}
\usepackage{lipsum}
\usepackage{CJK}
\usepackage{multirow}
\usepackage{lipsum}

\begin{document}
	

	\title{Experimental realization of  universal quantum gates and six-qubit entangled state using photonic quantum walk}
	
	
	\author{Kanad Sengupta }
	\affiliation{Dept of Instrumentation and Applied Physics, Indian Institute of Science, Bengaluru 560012, India}
	\author{S. P. Dinesh}
	\affiliation{Dept of Instrumentation and Applied Physics, Indian Institute of Science, Bengaluru 560012, India}
	\author{K.Muhammed Shafi}
	\affiliation{Dept of Instrumentation and Applied Physics, Indian Institute Science, Bengaluru 560012, India}
	\author{Soumya Asokan}
	\affiliation{Dept of Instrumentation and Applied Physics, Indian Institute Science, Bengaluru 560012, India}
	\author{C. M. Chandrashekar}
	\affiliation{Dept of Instrumentation and Applied Physics, Indian Institute Science, Bengaluru 560012, India}
	\affiliation{The Institute of Mathematical Sciences, C. I. T. Campus, Taramani, Chennai 600113, India}
	\affiliation{Homi Bhabha National Institute, Training School Complex, Anushakti Nagar, Mumbai 400094, India}
	
    \begin{abstract}
        For quantum computation using photons, performing deterministic quantum gate operations is a challenge due to the probabilistic nature of the photon-photon interaction. Encoding qubits in multiple degrees-of-freedom of photons and controlling operations between them is one of the promising ways to navigate the probabilistic behavior. Using single-photon discrete-time quantum walk in combination with polarization and path degrees-of-freedom, we experimentally demonstrate the realization of a universal set of quantum gates with high fidelity at room temperature. The deterministic realization of quantum gates through photonic quantum walk are characterized via quantum state tomography. For a three-qubit system using a single photon, the first qubit is encoded using polarization information, and the other two qubits are encoded using  path information, closely resembling a Galton-board setup. To generate a six-qubit Greenberger-Horne-Zeilinger state, entangled photon pairs are used to entangle the two three-qubit modules on which gate operations are performed. We also provide insights into the mapping of photonic quantum walk operations to quantum circuits and propose methods to resourcefully scale. This  demonstration marks a significant progress towards using quantum walks for quantum computing and provides a framework for using fewer photons in combination with different degrees-of-freedom of photon to scale the number of qubits.
    \end{abstract}

    \maketitle
        
	\section{Introduction}
        \hspace{0.75cm} Quantum computers promise to address problems that are beyond the reach of classical computation. By leveraging principles of quantum mechanics-including superposition, entanglement, interference, and uncertainty-they offer a powerful framework to model and explore complex quantum systems in nature, which are otherwise inaccessible to classical computing methods. Although numerous quantum algorithms have been developed and demonstrated in small scale on various quantum computing platforms\,\cite{BDC04,D2004,MA2024,KMNK2007, JF2008, SB2015, CI2017, FA2019, SG2019, NJ2022, JW2023}, the construction of a practical quantum computer with sufficient scale and capabilities remains an ongoing challenge. Multiple schemes for realizing qubits and performing quantum operations have been developed across different quantum computing platforms. Among them, the photonic platform employs diverse approaches such as discrete-variable, continuous-variable, measurement-based, and fusion-based methods\,\cite{RJ2001,PKT2005,HDW2009,SPHH2023}. 
        Since photons are naturally resistant to interactions with background electromagnetic fields, they can function efficiently at room temperature while maintaining stability and coherence during quantum operations. This makes them an ideal candidate for qubits compared to other systems despite the inherent probabilistic nature of the photon-photon interaction that has to be invoked to probabilistically implement quantum gate operations reducing the success probability of computational task\,\cite{KLM2001,JP2008,LR2013}.  However, probabilistic nature of the photon-photon interaction has also been an advantage and resulted in performing a large-scale complex computational task such as Boson sampling\,\cite{JBPW2012,MRSA2013,DEAR2019,HYZB2017,HS2021}. In addition, one can access various degrees-of-freedom of single photons to simultaneously encode multiple-qubits and experimentally control them to generate complex quantum states, demonstration of eighteen-qubit GHZ state generation using six-entangled photons is one such example\,\cite{XLW2018}. Any scheme demonstrating efficient realization of deterministic universal quantum gate operations among different degrees-of-freedom of photon  significantly enhances the prospects of the discrete-variable approach for photonic quantum computation. In this work we present a scheme to realize all universal set of quantum gate in combination of polarization and path degrees-of-freedom of single-photon using photonic quantum walk and report its experimental realization with high fidelity.

	Quantum walks, known in both continuous-time and discrete-time forms, have played a fundamental role in the development of quantum algorithms\,\cite{SKW03, CCD03, AA2003,XQH2024} and quantum simulation schemes\,\cite{OKA05, Str06, ECR07, MRL08, CL08, KRB10, Cha11, Cha13, MC16, MMC17, ASN20}. In particular, discrete-time one-dimensional quantum walks have been employed to engineer various high-dimensional quantum states, highlighting their versatility and potential, \cite{SCP10, TG2019}. These quantum walks have also been experimentally implemented using linear optical systems that control the polarization, path, and other degrees of freedom of single-photon states\,\cite{MAB10, PBR2010, AKV10, SAG12, YCH13, XTA2016, FYF2019}. 
Additionally, theoretical proposals report that quantum walks are also primitive for implementing quantum gates, thereby enabling universal quantum computation using both, continuous-time\,\cite{A09, NSM10, ADZ13} and discrete-time \cite{SPA2021, PSA2023} quantum walk variants. Among them, our mathematical model reported earlier, based on a discrete-time quantum walk,\cite{SPA2021,PSA2023}, offers a more tangible and logical foundation for the photonic quantum computing architecture in discrete-variable setting.  Therefore, in this work, we present a mapping of the quantum walk model to the photonic system involving polarization and path degree of freedom, present the architecture for implementing universal gate sets and experimentally demonstrate its validation.
    
    This polarization-path dual encoding facilitates the implementation of polarization-controlled operations and path-controlled operations, enabling quantum computations to be performed by emulating gates through unitary evolution. This approach offers a structured and physical framework for photonic quantum computation, making it a critical component for building quantum processors.
       
	 One of the main criteria for a system to be considered suitable for universal quantum computation is its ability to implement a universal set of quantum gates. The universal gate set typically includes single-qubit gates like phase(P) and Hadamard(H) gates, along with the two-qubit controlled-NOT(C-NOT) gate. The focus of this work is to harness the power of single-photon discrete-time quantum walk to develop a three-qubit module and demonstrate implementation of all the universal set of quantum gate operations on it and configure a six-qubit system using entangled photon pairs. For the experimental demonstration of gate operation on a three-qubit system, we use heralded single photons from the spontaneous parametric down conversion (SPDC) process and control their propagation across four path degrees of freedom. To demonstrate a six-qubit system, two three-qubit modules are combined using an entangled photon pair. A single photon's polarization and its interference along different paths achieve all multi-qubit gate operations in a three-qubit system. These are definite and not probabilistic in nature, making this a robust scheme. We report high fidelity for all universal quantum gate operations by performing quantum state tomography (QST). At six-qubit level, by controlling the entanglement between the two-photons polarization degrees of freedom and the path degrees in each three-qubit modules any six-qubit entangled state can be generated and we demonstrate generation of six-qubit Greenberger-Horne-Zeilinger (GHZ) state  which is characterized by the entanglement visibility. This work provides a scalable framework for photonic quantum computing  using lesser number of photons in combination with polarization and path degrees of freedom to increase the success rate of  the multi-qubit gate operations.

        \section{Results}

        \subsection{Photonic quantum walk and universal quantum gates.} The dynamics of a one-dimensional discrete-time quantum walk is defined on a configuration of Hilbert space composing of the coin space which is the internal degree of freedom of particle and position space, $\mathcal{H}_{w} = \mathcal{H}_{C}\otimes\mathcal{H}_{S}$\,\cite{CRR10, SE2012}. For a discrete-variable photonic quantum walk, the coin space is spanned by the horizontal and vertical polarization states of the photon, $\mathcal{H}_{C}=span$\{$\ket{H},\ket{V}$\}. A one-dimensional position space is spanned by  $\mathcal{H}_{S }=span$\{$\ket{-x},.......,\ket{-1},\ket{0},\ket{1},...,\ket{x}$\} representing the range of possible paths that a photon can take.
	
       The standard form of evolution operator for each step of quantum walk is defined using two sequential operations, $\hat{W}=[\hat{S}(\hat{C}(\theta)\otimes \mathbb{I})]$, first, is the unitary quantum coin operation acting only on the particles Hilbert space. A general form of quantum coin operation would be a $U(2)$ matrix consisting of 4 parameters, 
	\begin{equation}
		\hat{C}(\tau,\eta,\zeta,\theta) = e^{i\tau} \begin{bmatrix}
			e^{i\eta}\cos(\theta) & e^{i\zeta}\sin(\theta) \\
			-e^{-i\zeta}\sin(\theta) & e^{-i\eta}\cos(\theta) \\
		\end{bmatrix}.
	\end{equation}
        It is followed by the conditional shift operation that acts on the entire Hilbert space, 
	$$\hat{S} = \sum_{x}\Big[|H\rangle\langle H|\otimes|x-1\rangle\langle x|+|V\rangle\langle V|\otimes|x+1\rangle\langle x|\Big].$$
        If we consider the initial coin state to be $\ket{\psi}_{\text{C}} = \alpha|H\rangle +\beta|V\rangle$, then the initial state for the complete system will be $\ket{\Psi(t=0)} = \ket{\psi}_{\text{C}}\ket{x = 0}$. The state after $t$ steps will the be $|\Psi(t)\rangle=\hat{W}^{t}\ket{\Psi(t=0)}$. We experimentally achieve complete control on polarization state of a single photon using a combination of Quarter-Half-Quarter wave-plates (Q-H-Q), by neglecting the global phase of $ C(\tau,\eta,\zeta,\theta)$, i.e., $(\tau)$\,\cite{RN1990}. The path encoding and the polarization-dependent path encoding are accomplished by applying a beam-splitter (BS) and polarizing beam splitter (PBS), respectively. In different settings various realization of photonic quantum walks have already been experimentally demonstrated and what remains here would be to map the walk operations to universal quantum gate operations in polarization-path encoded qubits and demonstrate quantum computation using photonic quantum walk.

	The universal set of quantum gates for quantum computation consists of two single-qubit gates: the Phase gate (P) and the Hadamard gate (H), along with one two-qubit gate, the controlled-NOT gate (C-NOT). In mathematical form, it can be represented as follows,
	\[
	U = \left(\begin{bmatrix}1 & 0 \\ 0 & e^{i\pi/4}\end{bmatrix}, \frac{1}{\sqrt{2}}\begin{bmatrix}1 & 1 \\ 1 & -1\end{bmatrix}, \begin{bmatrix}1 & 0 & 0 & 0 \\ 0 & 1 & 0 & 0 \\ 0 & 0 & 0 & 1 \\ 0 & 0 & 1 & 0\end{bmatrix}\right).
	\] Combining these three gates we can achieve an optimal accuracy for all single and two qubit operation.
	
	\hspace{0.75cm}For the quantum computation model using discrete-time quantum walk a general form of shift operators and quantum coin operator is defined in     \,\cite {PSA2023,SPA2021}. The shift operations are configured to enable control on the shift of the particle to either left or to the right along with possibility of being in the same path conditioned  on the coin-state of the particle. In consolidated manner we can write the equation as                \begin{equation}
		\hat{S}_{\pm}^{k} = \sum_{l \in \mathbb{Z}} \Bigr[\ket{k}\bra{k} \otimes \ket{l \pm 1 \; mod \; m}\bra{l} + \sum_{i \neq j}^{\mu} \ket{i}\bra{i}\otimes\ket{l}\bra{l}\Bigr].
        \label{eq2}
	\end{equation}
    In Eq.\,\eqref{eq2} where $\mu$  is the total number of internal degrees of freedom and $\hat{S}^{0}_{+}$ or $\hat{S}^{0}_{-}$ dictates the shift of the particle to the right or left if the coin-state is $\ket{0}$ representing the polarization state of photon $\ket{H}$. Similarly, $\hat{S}^{1}_{+}$ and $\hat{S}^{1}_{-}$ act on $\ket{1}$ representing  the polarization state of photon $\ket{V}$, this type of shift operation can be achieved using a combination of  half-wave-plates (HWP) and PBS.  The $mod \; m$ comes for a closed graph dynamics of the quantum walk. The collection of operators $\{\hat{S}^{0}_{\pm},\hat{S}_{\pm}^{1},\hat{C}(\tau,\eta,\zeta,\theta),\mathbb{I}\}$ constitute a versatile set of operators representing the quantum walk. Moreover, this set of operators can be effectively employed to realize the universal set of quantum-gates within the discrete-time quantum walk framework.

    
    \subsection{Qubit encoding and universal gate implementation.} 
    
    Encoding qubits in the basis of the polarization degree and the path degree of freedom of photons forms the foundation of the photonic quantum computing approach presented here. For example, using the polarization state of a photon and a four-path basis for it, we can define a three-qubit system. While it is possible to extend the number of available paths for photons to increase the number of realizable qubits, this approach does not scale efficiently. Hence, we first present the encoding of two-qubit and three-qubit in the polarization and path basis and then addresses the challenges of scalability.
      \begin{table}[h!]
        \centering
        \caption{{\bf Q-H-Q angles for Pauli gates.}}
        \label{QHQ}
        \begin{tabular}{|c|c|c|}
        \hline
        \textbf{Gates} & \textbf{Q-H-Q angles} & \textbf{Operator} \\
        \hline
        $\sigma_{X}$ & \(\left(0, \frac{\pi}{4}, 0\right)\) & 
        \(\begin{pmatrix} 
        0 & 1 \\ 
        1 & 0 
        \end{pmatrix}\) \\
        \hline
        $\sigma_{Y}$ & \(\left(\frac{\pi}{2}, -\frac{\pi}{4}, 0\right)\) & 
        \(\begin{pmatrix} 
        0 & -i \\ 
        i & 0 
        \end{pmatrix}\) \\
        \hline
        $\sigma_{Z}$ & \(\left(0, \pi, \frac{\pi}{2}\right)\) & 
        \(\begin{pmatrix} 
        1 & 0 \\ 
        0 & -1 
        \end{pmatrix}\) \\
        \hline
        S\(\left(\frac{\pi}{2}\right)\) & \(\left(\frac{\pi}{4}, \frac{7\pi}{8}, \frac{5\pi}{4}\right)\) & 
        \(e^{i\pi/4}\begin{pmatrix} 
        1 & 0 \\ 
        0 & e^{i\pi/2} 
        \end{pmatrix}\) \\
        \hline
        P\(\left(\frac{\pi}{4}\right)\) & \(\left(\frac{3\pi}{4}, \frac{3\pi}{16}, \frac{3\pi}{4}\right)\) & 
        \(-e^{-i\pi/8}\begin{pmatrix} 
        1 & 0 \\ 
        0 & e^{i\pi/4} 
        \end{pmatrix}\) \\
        \hline
        \end{tabular}
        \end{table}
       \begin{figure}[H]
	\centering
	\includegraphics[width= 0.47\textwidth] {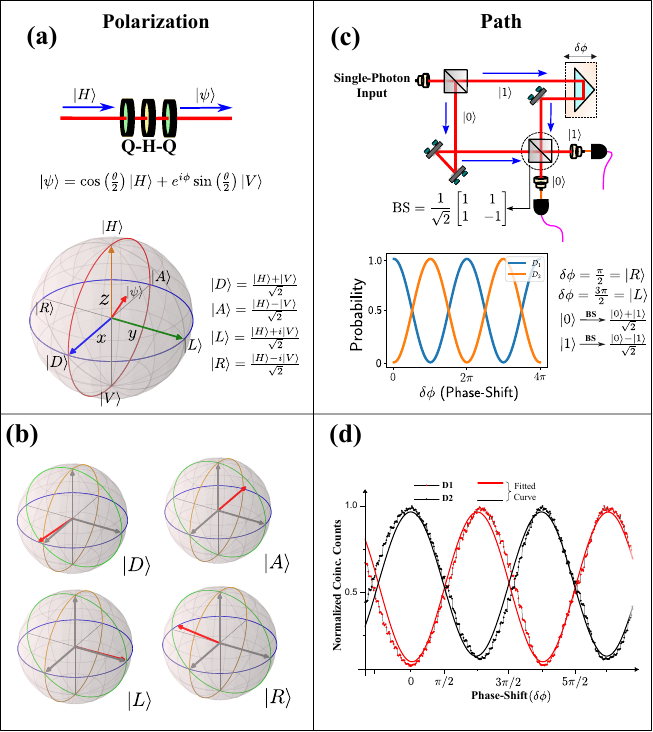}
	\caption{{\bf Qubit in polarization and path encoding.} In (a) and (c) illustration of encoding of polarization and path information of a single photon as qubits is shown. For polarization encoding, Q-H-Q plates are used to traverse the entire Bloch sphere. Similarly, for path encoding, phase adjustments in open and closed Mach-Zehnder interferometers enable access to all required orthogonal basis states. In (b) and (d) experimentally obtained data for single-photon polarization purity of $\ket{D}$, $\ket{A}$, $\ket{L}$, and $\ket{R}$ and the single photon interference visibility in path encoding indicating purity of path qubit is shown, respectively. D1 and D2 in the plots show the coincidence counts with the heralding photon, and solid red and black plot represent the numerical fit that is in agreement with the experimental data.}
	\label{Figure1}
    \end{figure}
   {\bf Polarization information as qubit :} A qubit on polarization degree of freedom can be written as,  
	\begin{equation}
		\label{eq:general_polarization}
		\ket{\psi} = \cos\left(\frac{\theta}{2}\right)\ket{H} + e^{i\phi}\sin\left(\frac{\theta}{2}\right)  \ket{V}.
	\end{equation}
Here varying the parameters, $\theta \in [0,\pi]$ and $\phi \in [0,2\pi)$ we can span all the points on the Bloch-sphere. This can be realized using a combination of two quarter wave-plates (QWP) and one HWP. The matrix structure of HWP and QWP derived from Jones calculus are, 
	\begin{equation}
		\begin{gathered}
			H(x) = \begin{pmatrix}
				\cos2x & \sin 2x \\
				\sin 2x & -\cos 2x
			\end{pmatrix} \\
			\\
			Q(x) = e^{-i\pi/4}\begin{pmatrix}
				{\cos^{2}x} + i\sin^{2}x & (1 - i)\sin{x}\cos{x} \\
				(1 - i)\sin{x}\cos{x} & i\cos^{2}x + \sin^{2}x
			\end{pmatrix}.
		\end{gathered}
	\end{equation}

    \onecolumngrid
    
     \begin{figure}[H]
    \centering
    {\includegraphics[width=1.0\textwidth]{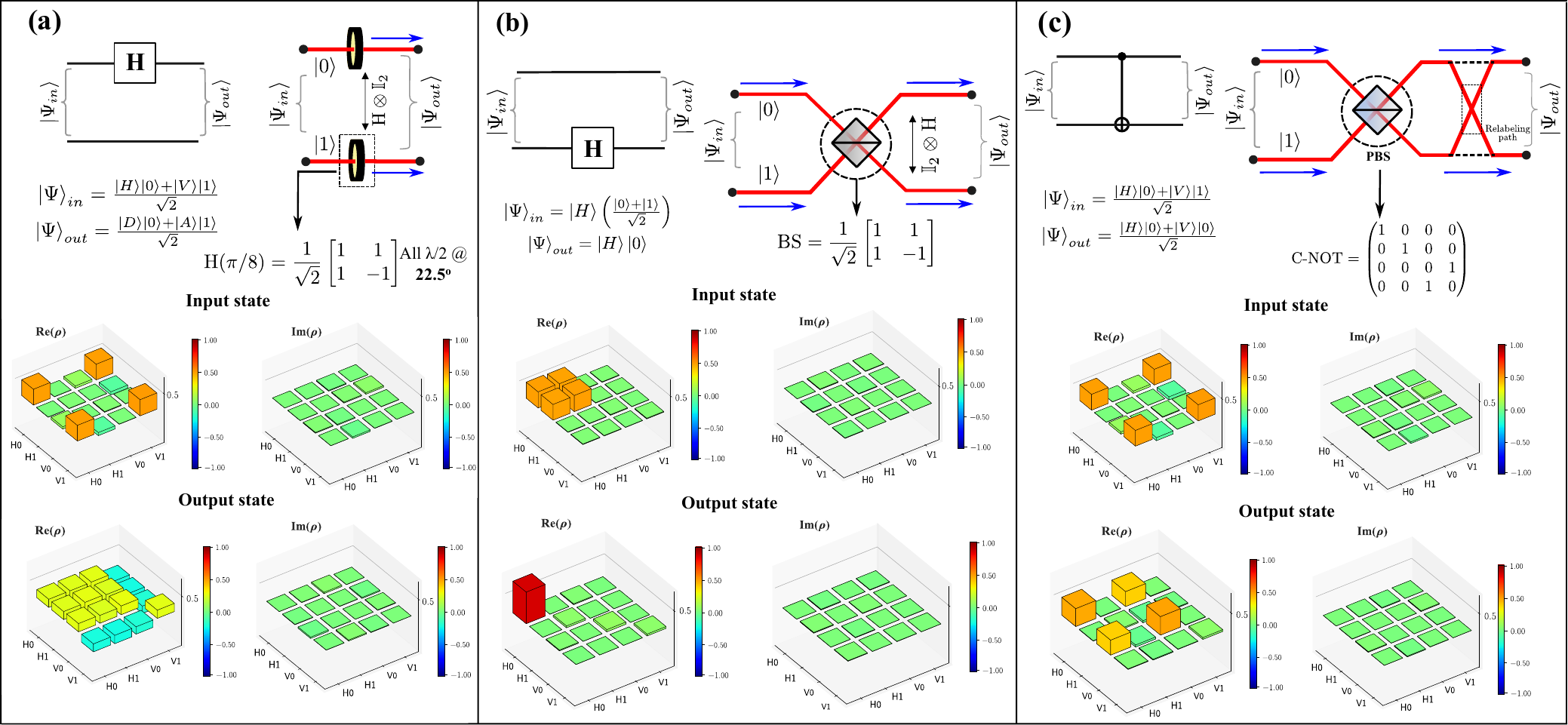}}
    \caption{{\bf Hadamard operation on the polarization and path qubit along with polarization controlled C-NOT operation on two-qubit state.} Illustration of implementation of the Hadamard gate on polarization and path qubits separately, as well as the two-qubit C-NOT gate, where polarization serves as the control, in panels (a), (b), and (c), respectively. Each subfigure presents the optical schematics of the gates, along with the experimentally obtained input and output states using quantum state tomography. For all the quantum states, both the real and imaginary parts of the tomography are shown. In panel (a), $\text{H}(\pi/8)$ represents a HWP set at an angle of $22.5^\circ$.}
    \label{figure_2}
    \end{figure}
    \twocolumngrid

Since the determinant value of the HWP and QWP are $-1$ and $+1$, respectively, the determinant value of the two QWP and one HWP will always be $-1$. The Pauli matrices \{$\sigma_{x}$, $\sigma_{y}$,  $\sigma_{z}$\} which forms the basis for single qubit gate operations also have a determinant value of $-1$. In Table\,\ref{QHQ} the  Q-H-Q angles to realize the Pauli matrices and phase, $\pi/4$ and $\pi/8$ are given. Using three wave-plates in different orders, Q-H-Q, Q-Q-H or H-Q-Q, with appropriate choice of angles, any arbitrary rotation on the Bloch sphere can be realized.

The set of Pauli matrices along with identity ($\mathds{I}_{2})$, $\{\mathds{I}_{2},\sigma_{x},\sigma_{y},\sigma_{z}\}$, form a basis for unitary matrices in the $(2\times2)$ space. By choosing the proper angles of Q-H-Q we can get all the Pauli matrices, enabling all the single-qubit operations on polarization qubits. In Fig.\ref{Figure1}(a), we demonstrate that any arbitrary polarization state \( |\psi\rangle \), which is an arbitrary linear combination of the states \( |H\rangle \) and \( |V\rangle \), can be generated from the horizontal polarization state \( |H\rangle \) using a combination of Q-H-Q plate.  Additionally, the experimental results for specific polarization states \( |L\rangle \), \( |R\rangle \), \( |A\rangle \), and \( |D\rangle \) are shown in Bloch sphere notation in Fig.\ref{Figure1}(b). 

    {\bf Path information as qubit :} Path encoding becomes most comprehensible through the lens of a Mach-Zehnder interferometer (MZI) designed for a single photon. The output quantum state of the MZI takes the form  $|\psi\rangle = \frac{(1+e^{i\delta\phi})}{2}\ket{0}+\frac{(1-e^{i\delta\phi})}{2}\ket{1}$, where $\ket{0}$ and $\ket{1}$ represent the path encoding of the photon achieved after the first beam-splitter. Here, $\delta\phi$ is the relative phase difference between $\ket{0}$ \& $\ket{1}$, in our experimental setup we use a nanometric precision piezo-stage in one of the arm to control $\delta\phi$. The output count of the interferometer varies as function of $\delta\phi$, counts in the $\ket{1}$ and $\ket{0}$ path follows $\cos^{2}(\frac{\delta\phi}{2})$ and $\sin^{2}(\frac{\delta\phi}{2})$, respectively. In Fig.\,\ref{Figure1}(c) we show the setup of MZI for a single photon interference and the probability of detecting a photon at the two output ports. In Fig.\,\ref{Figure1}(d) the measured visibility for the single-photon MZI is $V \approx 93.4\%$, is shown.

     In general, beam splitter matrix takes the form 
     \begin{equation}
         U_{\text{BS}} =\begin{bmatrix}
            \sqrt{T}  & -e^{-i\phi}\sqrt{R} \\
            e^{i\phi}\sqrt{R} & \sqrt{T} 
        \end{bmatrix},
     \end{equation}
     if we consider $\text{transmitted}(T) = \text{reflected}(R) = 1/2$ and $\phi = 0$, then we get back the general 50:50 beam-splitter matrix, $U_{\text{BS}}$ will take the form 
    $U_{\text{BS}}(\phi = 0) =\frac{1}{\sqrt{2}}\begin{bmatrix}
    1  & -1 \\
    1  &  \,\,1
    \end{bmatrix}.$ Since, BS as a Hadamard operation for our purpose, we always consider an extra phase of $\pi$ to the $\ket{1}$ arm, so that the final unitary transformation matrix exactly resembles Hadamard transformation\,\cite{MD2022,GA2001}.

  From the perspective of quantum information encoding, manipulating $\delta\phi$ enables the preparation of all fundamental states: $\ket{+}$, $\ket{-}$, $\ket{L}$, and $\ket{R}$. This highlights the versatility of path encoding and its potential for information manipulation in the quantum domain. In our context, path encoding initially disregards polarization information.

\onecolumngrid

    \begin{figure}[H]
    \centering
    {\includegraphics[width = 1.0\textwidth]{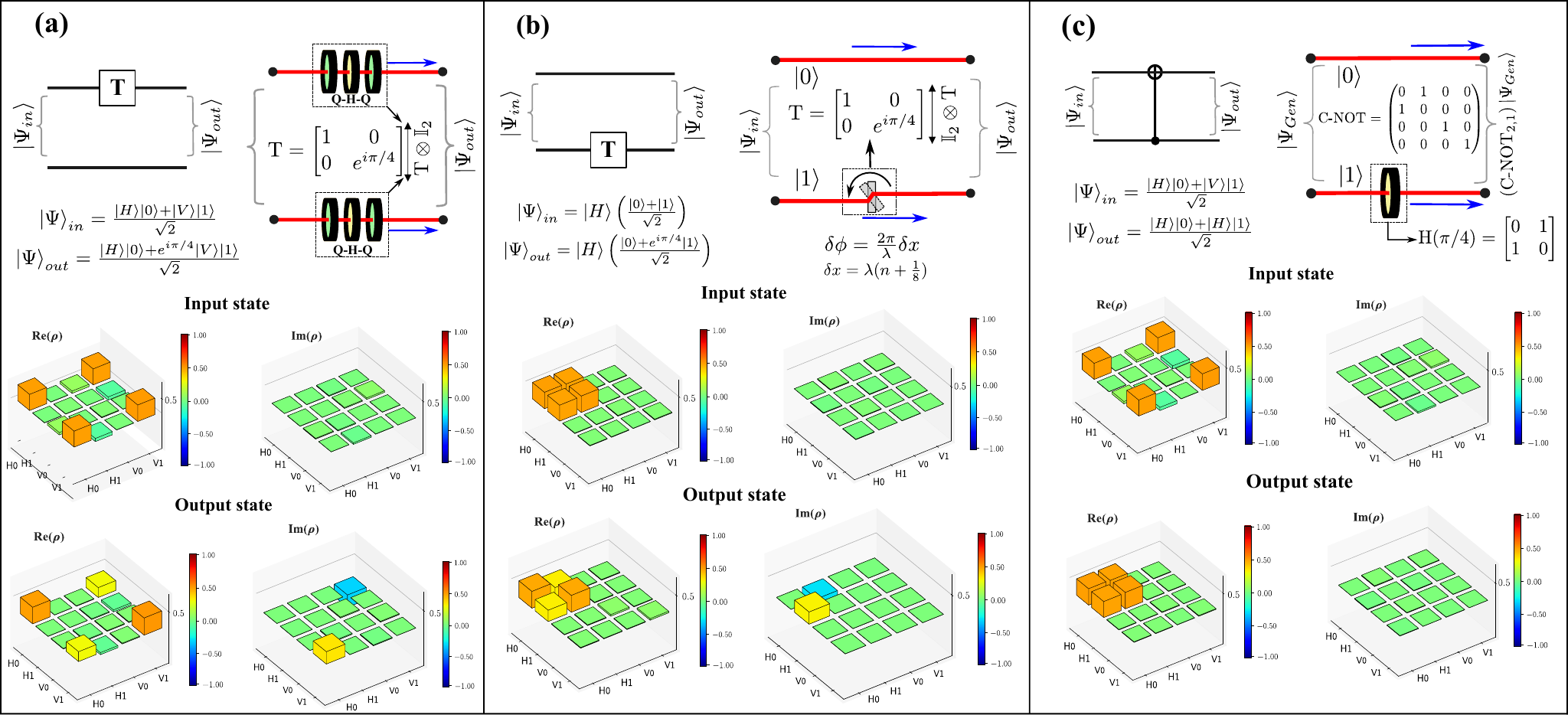}}
    \caption{{\bf Phase gate for polarization and path qubit along with path controlled C-NOT operation of two-qubit state.}  Illustration of implementation of phase gate on polarization and path qubits separately, as well as the two-qubit C-NOT gate, where the path serves as the control, in panels (a), (b), and (c), respectively. In panel (a), $\text{Q-H-Q}$ denotes a combination of wave plates, aligned at specific angles to introduce a $\pi/8$ phase difference between $\ket{H}$ and $\ket{V}$, with the global phase defined as previously described. For the phase gate in the path degree of freedom, an additional phase delay is introduced using a glass beam splitter controlled by a piezoelectric stage. Finally, in panel (c) $\text{H}(\pi/4)$ represents a half-wave plate set at an angle of $45^\circ$, which will convert $\ket{H/V}$ and $\ket{V/H}$, for path $\ket{1}$, which serves as C-NOT operation. Each subfigure presents the optical schematics of the gates, along with the experimentally obtained input and output states using quantum state tomography. For all the quantum states, both the real and imaginary parts of the tomography are shown.}
    \label{figure_3}
    \end{figure}

    \twocolumngrid	

  To achieve a comprehensive description of polarization-path encoding, we must account for the polarization of each path. This results in a Hilbert space denoted as $\mathcal{H}_{\text{P-P}} = \mathcal{H}_{\text{Pol.}} \otimes \mathcal{H}_{\text{Path}}$, encompassing a total of four modes for a single photon, two polarization states and two path states, effectively representing the polarization-path quantum state.
	The general two-qubit state can be expressed as:
	\begin{equation}
		\ket{\Psi}_{\textrm{Gen}} = \alpha\ket{H}\ket{0} + \beta\ket{H}\ket{1} + \gamma\ket{V}\ket{0} + \delta\ket{V}\ket{1},
        \label{EqG}
	\end{equation}
	where $\{\alpha,\beta,\gamma,\delta\} \in \mathcal{C}$ and $|\alpha|^{2}+|\beta|^{2}+|\gamma|^{2}+|\delta|^{2} = 1$. A general two-qubit state preparation is discussed in supplementary section. The computational basis takes the form - $\ket{H}\ket{0} = \ket{00}$, $\ket{H}\ket{1} = \ket{01}$, $\ket{V}\ket{0}=\ket{10}$ and $\ket{V}\ket{1} = \ket{11}$.\\

To prepare a two-qubit entangled state in polarization-path encoding, we use a PBS to split the polarization information of the photon to different paths. If the input state in polarization and path combination is  $\ket{\psi}_{\text{in}} = (\alpha\ket{H}+\beta\ket{V})\ket{0}_{\text{in}}$, where the $\ket{0}_{\text{in}}$ represents input path of PBS, then the output state becomes $\ket{\psi}_{\text{out}} = \alpha\ket{H}\ket{0}_{\text{out}}+\beta\ket{V}\ket{1}_{\text{out}}$, a polarization-path entangled state, $\ket{1}$ represents the other arm of the PBS. \\

    {\it One- and two-qubit gate implementation:}  For realization of two-qubit state in polarization-path combination, heralded single photons of 810 nm, obtained via SPDC process using a PPKTP crystal, were used. To perform a QST and reconstruct the density matrix, an additional path interference are involved. To experimentally obtain the gate fidelity for all universal gates, we prepared different initial states on which the specific gate action are prominent and involves the coherence between polarization-path encoding. For $\{H \otimes \mathbb{I}_{2}, \text{C-NOT}(1,2), P \otimes \mathbb{I}_{2}, \text{C-NOT}(2,1)\}$, the initial state was prepared as $|\Psi\rangle_{\text{in}} = \frac{\ket{H}\ket{0} + \ket{V}\ket{1}}{\sqrt{2}}$, 
    and the gate operations were implemented, followed by fidelity calculations. For the remaining two gates $\{\mathbb{I}_{2} \otimes H, \mathbb{I}_{2} \otimes P\}$, we considered the initial state to be $\ket{H} \big(\frac{\ket{0} + \ket{1}}{\sqrt{2}}\big)$. Starting from state preparation to QST, on an average of ten gate operations in the form of standard circuit model are executed. In Fig.\,\ref{figure_2}(a)-(c) illustrate the circuit model, optical schematic, along with the input and
experimentally obtained output states using quantum state tomography for the Hadamard gate on polarization, path qubits, and the C-NOT gate having polarization as control, applicable to the general quantum state in Eq.(\ref{EqG}). Similarly, Fig.\,\ref{figure_3}(a)-(c) illustrate the circuit model, optical schematic, and along with the experimentally obtained states using QST for the phase($\pi/8$) gate on polarization and path qubits, and the C-NOT gate having path as control, applicable to the general quantum state in Eq.\,(\ref{EqG}).

For all the quantum states, both the real and imaginary parts of the tomography are shown. In Table\,\ref{tab:table1} fidelity of all the one- and two-qubit quantum gate operation on two-qubit state composing of polarization and path degree-of-freedom obtained after performing QST is reported. The role of visibility of each interferometer in multi-path interferences configuration on the final fidelity and the measure to improve the reported fidelity is discussed in {\bf Methods} section. In this quantum walk framework, it is also possible to implement all the universal gate set using path–path encoded qubits. For example, the two-step quantum walk evolution will give access to four spatial modes: ${\ket{00}, \ket{01}, \ket{10}, \ket{11}}$, which naturally encode two-qubit in the path degree of freedom. The implementation of single-qubit gates such as the Hadamard and Phase gates in this encoding is identical to the one descripbed in the path degree of freedom in the polarization-path ecoding scheme. Moreover, the CNOT gate between the path-encoded qubits can be realized more easily by appropriately rerouting or relabeling the paths, making the implementation two-qubit gate straight forward and efficient.\\
  
\begin{figure}[H]
	\centering
	\includegraphics[width= 0.47\textwidth]{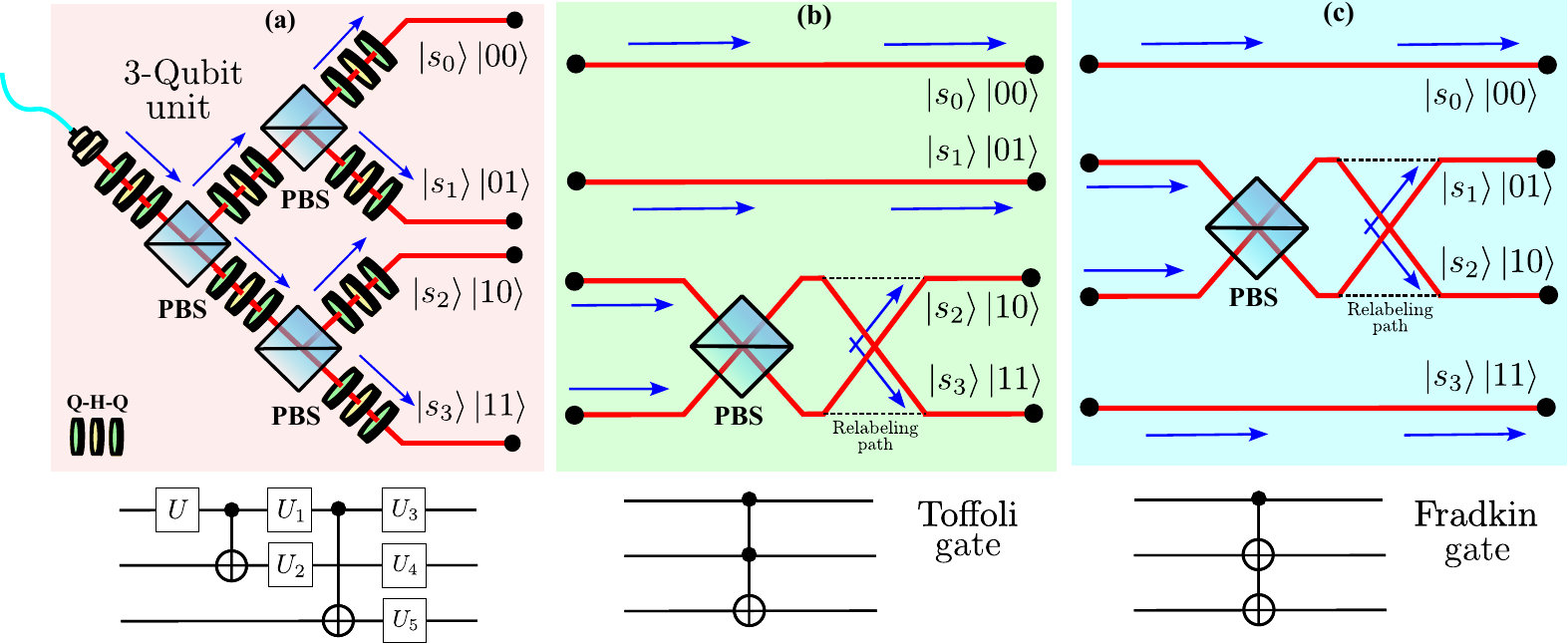}
	\caption{{\bf Three-qubit system in polarization-path-path encoding.} In (a) the schematic of the circuit and optical scheme to preparing a polarization-path-path encoded three-qubit system is shown. The (b) \& (c) depict the circuit as well as optical scheme to implement the Toffoli and Fradkin gate on the three-qubit state.}
  	\label{fig:3-qubit s}
\end{figure}

{\bf Three and six-qubit system:} For a three-qubit system, the polarization degree of freedom of a photon represents the first qubit, while the four path degrees of freedom represent the second and third qubits, respectively. The Hilbert space is given by $\mathcal{H}_{\text{Pol.}} \otimes \mathcal{H}_{\text{Path}_1} \otimes \mathcal{H}_{\text{Path}_2}$. The implementation of Toffoli and Fradkin gate operations on this three-qubit system in a standard circuit model, along with their equivalent optical schemes in the polarization-path-path encoded three-qubit system, is illustrated in Fig.\,\ref{fig:3-qubit s}. Additionally, it is worth noting that realizing the Hadamard operation on a path-encoded qubit involves interferometry, and implementing the C-NOT gate, which is typically challenging in other setups, is simplified in this scheme by swapping the paths, as shown in the Appendix, Fig.\,\ref{fig:3qubit_all}. 
     \begin{table}[h!]
        \caption{\label{tab:table1}Fidelity of two-qubit gate operation.}
        \begin{ruledtabular}
            \begin{tabular}{c|c|c|c|c|c|c}
                \textrm{\textbf{Gate}} &
                $\textrm{H}\otimes \mathbb{I}_{2}$ &
                $\textrm{T}\otimes\mathbb{I}_{2}$ &
                $\textrm{C-NOT}_{1,2}$ &
                $\mathbb{I}_{2}\otimes\textrm{H}$ &
                $\mathbb{I}_{2}\otimes\textrm{T}$ &
                $\textrm{C-NOT}_{2,1}$ \\
                \hline
                F & 0.97(8)  & 0.96(7) & 0.91(2) & 0.92(9) & 0.94(7) & 0.97(4)\\
            \end{tabular}
        \end{ruledtabular}
    \end{table}

To calculate the fidelities  of the gate operations, starting from the three-qubit general state preparation to performing  quantum state tomography, maximum of  twenty eight gate operations are executed. Since the fidelities of Hadamard operations are same as given in Table\,\ref{tab:table1}, only C-NOT operations for different combination of qubits are given. We can note that the fidelity  of C-NOT gate between two path qubits is very high when compared to C-NOT between the polarization and path qubits.  This is because to use of only swapping of path between path qubits to implement C-NOT gate. This allows for efficient handling multi-qubit operations using path basis.

\begin{figure}[H]
		\centering
		\includegraphics[width= 0.95\textwidth]{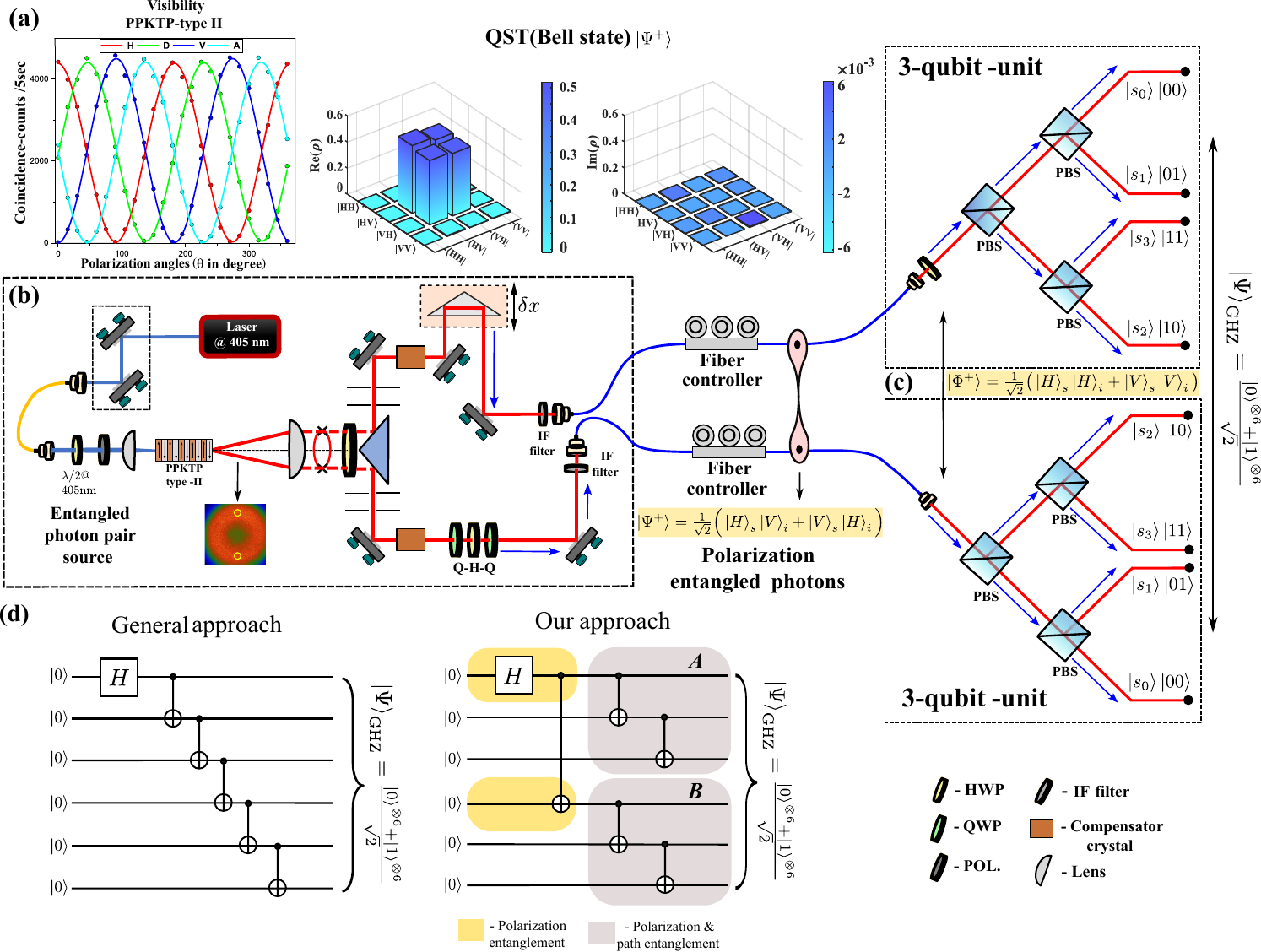}
		\caption{{\bf Six-qubit system using two photons in polarization-path-path degrees-of-freedom.} In (a) the visibility observed in the four different basis and the QST for the polarization entangled state $\ket{\Psi^{+}}$ from the type-II SPDC process illustrated in subfigure (b) is shown. The entangled photons are coupled to two fiber paddle controllers which has an independent wave plates that will alter the polarization of the transmitted light in a single mode fiber before being directed to two separate three-qubit modules (c), as shown in the right. By utilizing the entanglement between two photons, we connect the distinct three-qubit units and perform the necessary gates on each unit as required. For the preparation of the six-qubit GHZ state, we apply two consecutive C-NOT gates, with polarization serving as the control in each three-qubit unit. The schematic of the circuit for our approach to prepare the six-qubit GHZ state and the  general GHZ state preparation is shown in (d).$\ket{s_{i}}$ in represents the polarization information of each path.}
		\label{fig:6qubit}
	\end{figure}

 One of the method to realize six-qubit system is illustrated in Fig.\,\ref{fig:6qubit}.   A six-qubit system is sourced by the, entangled photon pair  given by  $\ket{\Psi} = \frac{1}{\sqrt{2}}(\ket{H}_{s}\ket{V}_{i} + \ket{V}_{s}\ket{H}_{i})$ generated  from the  SPDC process using type-II PPKTP crystal.  Each photon from the entangled pair are used as an input  to two separate three-qubit systems.  Various operations on the six-qubit system can be configured first by controlling entanglement between the two photons and these two photons are  sourced to the two three-qubit systems where gate operations are performed as discussed for three-qubit system earlier.  For example, we can easily convert  entangled state to  the state  $\ket{\Psi^{'}} = \frac{1}{\sqrt{2}}(\ket{H}_{s}\ket{H}_{i} + \ket{V}_{s}\ket{V}_{i})$ using HWP in one of the the arm (signal arm). By placing a  PBS in the two separate three qubit units we can generate the six-qubit GHZ state 
 $$\ket{\Psi}_{GHZ} = \frac{1}{\sqrt{2}}(\ket{0}^{\otimes 6} + \ket{1}^{\otimes 6}).$$

 Though all the six-qubits in the system are not directly connected to one another, through entangled photon pairs they can be connected and engineered to control a six-qubit system as shown in Fig.\,\ref{fig:6qubit}(d). To assess the quality of the prepared GHZ state, we used an entanglement witness measurement in the path states associated with specific polarizations. Due to the C-NOT gate operations, we performed visibility checks between the \(\ket{00}\) and \(\ket{10}\) paths of the two three-qubit units. Similarly, visibility checks were conducted between the \(\ket{01}\) and \(\ket{11}\) paths of the two three-qubit units. The visibility in the H and D bases for these paths was measured to be \(98.3\%\) and \(95.1\%\), respectively. From the structure of the six-qubit GHZ state and the preparation method illustrated in Fig.\,\ref{fig:6qubit}, it is evident that coincidences occur between the \(\ket{00}\) and \(\ket{11}\) paths of the three-qubit units for \(\ket{H}_{s}\ket{00}\ket{H}_{i}\ket{00}\) and \(\ket{V}_{s}\ket{11}\ket{V}_{i}\ket{11}\), respectively. The number of accidental coincidences between the other paths of two three-qubit units are low as predicted by our way of preparation. By performing these coincidence measurements and visibility checks, which serve as entanglement witnesses, we confirm the successful preparation of the GHZ state. \\

 {\bf Scaling :} Number of qubits can be scaled by using multiple four path basis in a tensor network configuration of  two-qubits and entangling them with a single photon  as as given in the mathematical model of the  scheme\,\cite{SPA2021, PSA2023}. However,  single photon in multiple  network of two-qubit position basis will results in loss of fidelity.  As demonstrated for six-qubit here,  further scaling  with a combination of six-qubit systems can be  explored by entangling other degree's of freedom of photons and involving more photons. However, involving more photons makes the connection between the each modules of smaller qubits probabilistic but overall efficiency becomes better than the standard approach of using single photon in dual rail encoding to represent each qubits. 
 \onecolumngrid	

\twocolumngrid
 The discrete-time quantum walk framework reported here can be readily implemented on a photonic integrated circuits (PICs), provided polarization control within the circuit is achievable and promising results are being reported in that direction \cite{JW2023}.  The reported fidelities for various quantum gates using our scheme in bulk optics can be improved in PICs which offer advantages over bulk optics, including better scalability by mitigating issues like alignment and sustained stability. At the same time, bulk optics provides flexibility and ease of reconfiguration, which are valuable for experimental prototyping and validation.

  \begin{figure}[H]
        \centering
	\includegraphics[width= 0.47\textwidth]{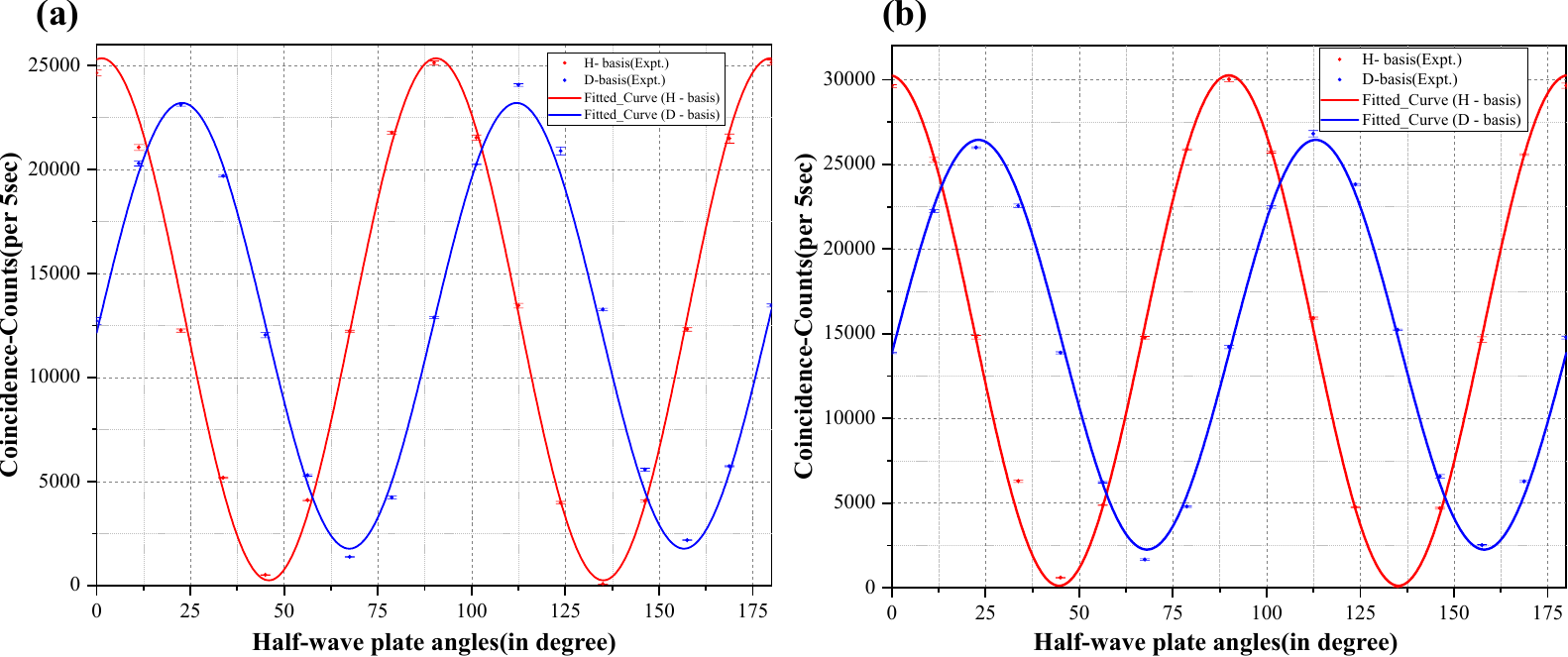}
	\caption{{\bf Entanglement witness of the GHZ state.} The plots (a) and (b) shows the coincidence between  arms $\ket{00}$, $\ket{00}$ and $\ket{11}$, $\ket{11}$ for H and D basis respectively, shown in Fig\,\ref{fig:6qubit}. The red and blue dots are the data points measured experimentally and solid blue, red lines are the numerical fits. The obtained coincidence visibility for H \& D basis is $98.3\%$ and $95.1\%$. This entanglement witness measurement in the path states which carry a specific polarization can be inferred to asses the quality of the prepared GHZ state.}
    \label{ENTG_wit}
    \end{figure}

\section{Conclusion}
In this work we have report the experimental demonstration of realizing universal set of quantum gates using photonic quantum walk. In the scheme, multiple qubits were encoded using polarization and paths degree of freedom of single photon and universal quantum gates were implemented with 100\% success probability, and high fidelity.  The fidelity of each gate operations were characterized by performing quantum state tomography in room temperature setting. For a two-qubit and three-qubit system, the first qubit is encoded with polarization state of photon and  path information was used for encoding other qubits. For the six-qubit system an entangled photon pairs from SPDC process was used to source two three-qubit modules and realize full six-qubit  GHZ state. In the scheme  only one photon is involved  to realize two-qubit and three-qubit system. Not invoking  direct photon-photon interaction has resulted in very high fidelity quantum gate operations.  This also makes  all the gates operation a definite realization. Fidelities directly accounts for experimental inaccuracies involved in the process. We have demonstrated that some of the two-qubit gates, C-NOT between path qubits,  are easily realizable in the scheme compared to other approaches. This quantum walk based quantum computing scheme can be extended to other photonic degrees of freedom—such as frequency, time-bin, and orbital angular momentum—using hybrid encoding strategies. This flexibility enables scalable and high-dimensional quantum photonic architectures. This experimental demonstration reporting the long standing theoretical proposal of realizing quantum gates using quantum walks indicates a promising  progress towards using photonic quantum walk for quantum computing. It also provides a general framework for photonic quantum computing in room temperature setting using lesser number of photons in combination with path and other degree of freedom of photons.  

    \section{Methods}
    To generate heralded single-photons and entangled photon pairs, type-II spontaneous parametric down-conversion (SPDC) process using PPKTP crystal was used. A 10-mm-long PPKTP crystal was pumped using a continuous-wave (CW) diode laser at 405 nm (Toptica TopMode laser @ 405 nm). The astigmatism of the pump beam was corrected by coupling it through a single-mode fiber (SMF), and the collimated output was horizontally polarized using a wire-grid polarizer Fig.\,\ref{fig:6qubit}(b). A plano-convex lens (L1) with a focal length of $17.5 , \mathrm{cm}$ focused the beam at the center of the crystal, which was maintained at a temperature of $32.3^\circ \mathrm{C}$ to achieve degenerate  SPDC at 810 nm. An SPDC cone angle of $0.83^\circ$ was chosen to optimize photon collection. The down-converted photons were collimated using another plano-convex lens (L2) with a focal length of $10, \mathrm{cm}$. Applying a PBS in th path of down-converted photons one of photon (signal) in photo pair of specific polarization is chosen and other photon (idler) used for heralding purpose. At a pump power of 15 mW, the photon generation rate is $6.93×10^{5}$ counts/s, with a heralding efficiency of 22.3\% after coupling to a single-mode fiber. We have used the single photon counting module (SPCM) as SPCM-800-44-FC, Excelitas, whose detection efficiency at $810$ nm is approximately 63\%. 

    For entangled photon pair generation, a prism mirror was used to separate the down-converted rings into two halves. Two diametrically opposite points were then carefully collected from each half of the ring. A half-wave plate (HWP) set at $45^\circ$ and 5-mm KTP crystals for temporal compensation were introduced in each path, as shown in Fig.~\ref{fig:6qubit}(b). Residual pump photons were removed using bandpass filters ($810 \, \mathrm{nm} \pm 10 \, \mathrm{nm}$), and polarization integrity was maintained using fiber polarization controllers (FPCs) \cite{SHMH2016,MCF2005,SK2023}. This configuration produced polarization-entangled photon pairs at a rate of approximately $16400, \mathrm{pairs/s/mW}$. Quantum state tomography (QST) indicated a fidelity of $98.12\%$ with the ideal Bell state. At a pump power of $0.15 , \mathrm{mW}$, accidental counts were reduced to 16 counts per second. The measured visibilities in the H/V and A/D bases were $98.6\%$ and $98.1\%$, respectively, with a Bell-CHSH inequality violation of $S = 2.77 (\pm 0.001)$.

     Using the heralded single-photon source, a polarization-path entangled state is prepared, and all universal gates are implemented. The input and output states are characterized via tomography, as detailed in the supplementary material. Building on the polarization-path encoded state, the two paths are further split to realize polarization-path-path encoded state, enabling a three-qubit system. Finally, for six-qubit GHZ state preparation, we utilize the entangled photon pairs and two separate three-qubit units, each consisting of polarization-path-path encoded states.

    For two-qubit state tomography requiring external (additional) path interference from the output state, the beam splitter used for interference is not exactly 50:50(BS005) for both S- and P-polarizations. This discrepancy significantly impacts the fidelity of all the universal quantum gates we considered.

    Path-encoded qubits are sensitive to phase instability. Although we did not employ active stabilization methods, we minimized the phase fluctuations between the two arms of interferometer by using passive stabilization techniques. The interferometer was fully enclosed using a custom-built isolation cover and mounted within cage assemblies to suppress air currents and enhance environmental isolation. To minimize path length variations caused by temperature changes, precision mirror mounts with low thermal expansion coefficients were employed, effectively reducing phase fluctuations. Additionally, motorized wave-plates were incorporated to avoid repeatedly opening and closing the enclosure, thereby maintaining consistent environmental conditions. Optical fibers used in the setup were jacketed and thermally insulated to further suppress phase noise arising from temperature fluctuations.
    \color{black}
    For specific cases, the performance of the Hadamard gate on the polarization qubit can be improved by using a motorized half-wave plate with high angular precision. Similarly, the Hadamard gate on the path qubit inherently relies on path interference, which slightly lowers its fidelity during QST. For the polarization-controlled C-NOT gate, we used a PBS with an extinction ratio of 1000:1. Enhancing this extinction ratio would lead to better results. Although the phase gate for polarization involves three wave plates in each arm, reducing the number of wave plates in one path and using high-precision rotation mounts can improve its performance. The phase gate in the path qubit requires nanometric piezo-controllers and high stability, which, if utilized, can enhance its accuracy. A similar explanation applies to the C-NOT gate, where the path acts as the control, and precise wave plate adjustments are essential for optimal operation.

      \section{Appendix}
        All the three-qubit universal gates optical schematics are listed, below. It is worth noting in C-NOT gate in path-path qubits are just swapping the path or relabeling, it provides a great advantage with respect to other quantum computing architecture. 
\onecolumngrid	

	   \begin{figure}[H]
		  \centering
		  \includegraphics[width = 0.85\textwidth]{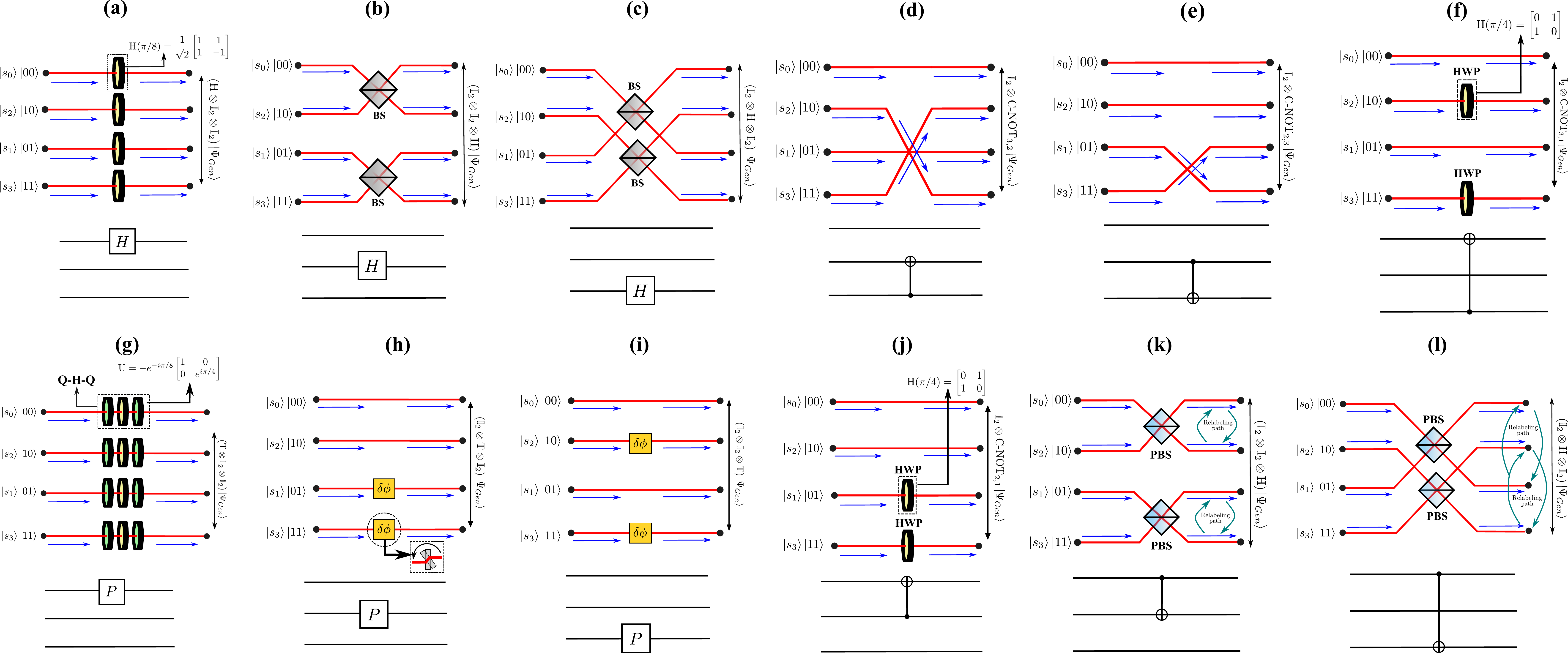}
		  \caption{{\bf Gate operations on three-qubit system in polarization-path-path configuration.}Schematic illustration of all the three-qubit gate operations for polarization-path-path encoded quantum states. As mentioned earlier, the four paths are labeled as $\ket{00}$, $\ket{01}$, $\ket{10}$, and $\ket{11}$, arranged from top to bottom in the optical diagrams. Diagrams (a), (b), and (c) depict the Hadamard gate applied to polarization, the first path qubit, and the second path qubit, respectively. In (d), a C-NOT gate is shown where the first path qubit serves as the control for the second, implemented by relabeling the $\ket{10}$ and $\ket{11}$ paths. Diagram (e) shows the second path qubit controlling the first qubit, achieved by swapping the $\ket{01}$ and $\ket{11}$ paths. In (f), a polarization-controlled C-NOT gate is implemented for the second path qubit. The phase-gate ($\pi/8$) operation is illustrated in (g), (h), and (i) for polarization, the first path qubit, and the second path qubit, respectively. Diagram (j) shows a polarization-controlled C-NOT gate for the first path qubit, while (k) demonstrates the implementation of such a gate for the first path qubit. Finally, (l) represents a polarization-controlled C-NOT operation for the second path qubit.
}
		\label{fig:3qubit_all}
	\end{figure}

\twocolumngrid


\vskip 0.4in 
	\noindent
    {\bf Acknowledgement }\\
	We would like to thank Dr. Prateek Chawla and Dr. Sujai Matta for useful discussions on some of the schemes for optical realization and support in the experiment.\\

\vskip 0.2in
\noindent
{\bf Author contributions} \\
CMC  conceived the idea. All authors contributed towards developing the experimental setup, performing experiments and collecting the data. KS and CMC accomplished theoretical calculations and data analysis. The manuscript was written through contributions from all authors.\\

\vskip 0.2in
\noindent
{\bf Funding}\\
We acknowledge the support from the Office of Principal Scientific Advisor to the Government of India, project no. Prn.SA/QSim/2020.

\vskip 0.2in
\noindent
{\bf Competing interests}\\
The authors declare no competing interests.

\vskip 0.2in
\noindent
{\bf Data Availability}
The data that support the findings of this study are available from the corresponding author upon reasonable request.

\end{document}